\documentclass[prb,superscriptaddress,showpacs,a4paper,twocolumn,floatfix]{revtex4}

\usepackage{amsmath} 
\usepackage{graphics} 
\usepackage{dcolumn} 
\usepackage{epsfig} 

\begin{document}

\title{Substrate dependent bonding distances of PTCDA \\ -- A comparative XSW
  study on Cu(111) and Ag(111)}

\author{A. Gerlach} 
\author{S. Sellner}
\author{F. Schreiber}
\email[corresponding author:]{frank.schreiber@uni-tuebingen.de}

\affiliation{Institut f\"ur Angewandte Physik, Universit\"at T\"ubingen, Auf
  der Morgenstelle 10, 72076 T\"ubingen, Germany} 

\author{N. Koch}

\affiliation{Institut f\"ur Physik, Humboldt-Universit\"at zu Berlin,
  Newtonstr. 15, 12489 Berlin, Germany} 

\author{J. Zegenhagen}

\affiliation{European Synchrotron Radiation Facility, 6 Rue Jules Horowitz,
  BP 220, 38043 Grenoble Cedex 9, France}

\date{\today}

\begin{abstract} 
  We study the adsorption geometry of
  3,4,9,10-perylene-tetracarboxylic-dianhydride (PTCDA) on Ag(111) and Cu(111)
  using X-ray standing waves. The element-specific analysis shows that the
  carbon core of the molecule adsorbs in a planar configuration, whereas the
  oxygen atoms experience a non-trivial and substrate dependent distortion. On
  copper (silver) the carbon rings resides $2.66\,$\AA{} ($2.86\,$\AA) above
  the substrate. In contrast to the conformation on Ag(111), where the
  carboxylic oxygen atoms are bent towards the surface, we find that on
  Cu(111) all oxygen atoms are above the carbon plane at $2.73\,$\AA{} and
  $2.89\,$\AA, respectively.
\end{abstract}

\maketitle

\section{Introduction}
\label{sec:intro}

In recent years the adsorption of $\pi$-conjugated molecules on various
surfaces has received significant attention.\cite{eremtchenko_n03,witte_jmr04}
Still, one of the fundamental parameters in the adsorption process, the
bonding distance of the first layer to the substrate, is largely unknown for
most organic adsorbate systems.  Measuring this quantity with the required
precision poses a serious experimental challenge which requires specialized
methods.  As the adsorbate distance of the molecules is closely related to the
character of the bond, it is also highly desirable to combine structural and
spectroscopic techniques in the experiment. A suitable approach could reveal
correlations between the bonding distances and the electronic properties of
the adsorbate complex on different surfaces, in particular when the nature of
the bonding is controversial.

\begin{figure}[htbp]
  \centering
  \begin{minipage}{\columnwidth}
  \includegraphics[height=4cm]{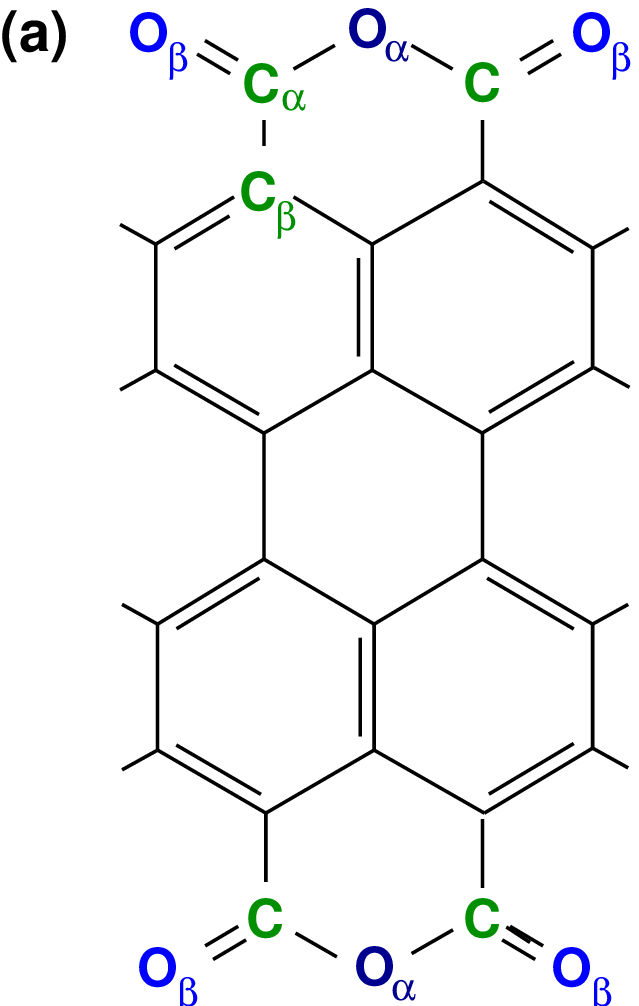}%
  \hspace{3mm}
  \includegraphics[height=4cm]{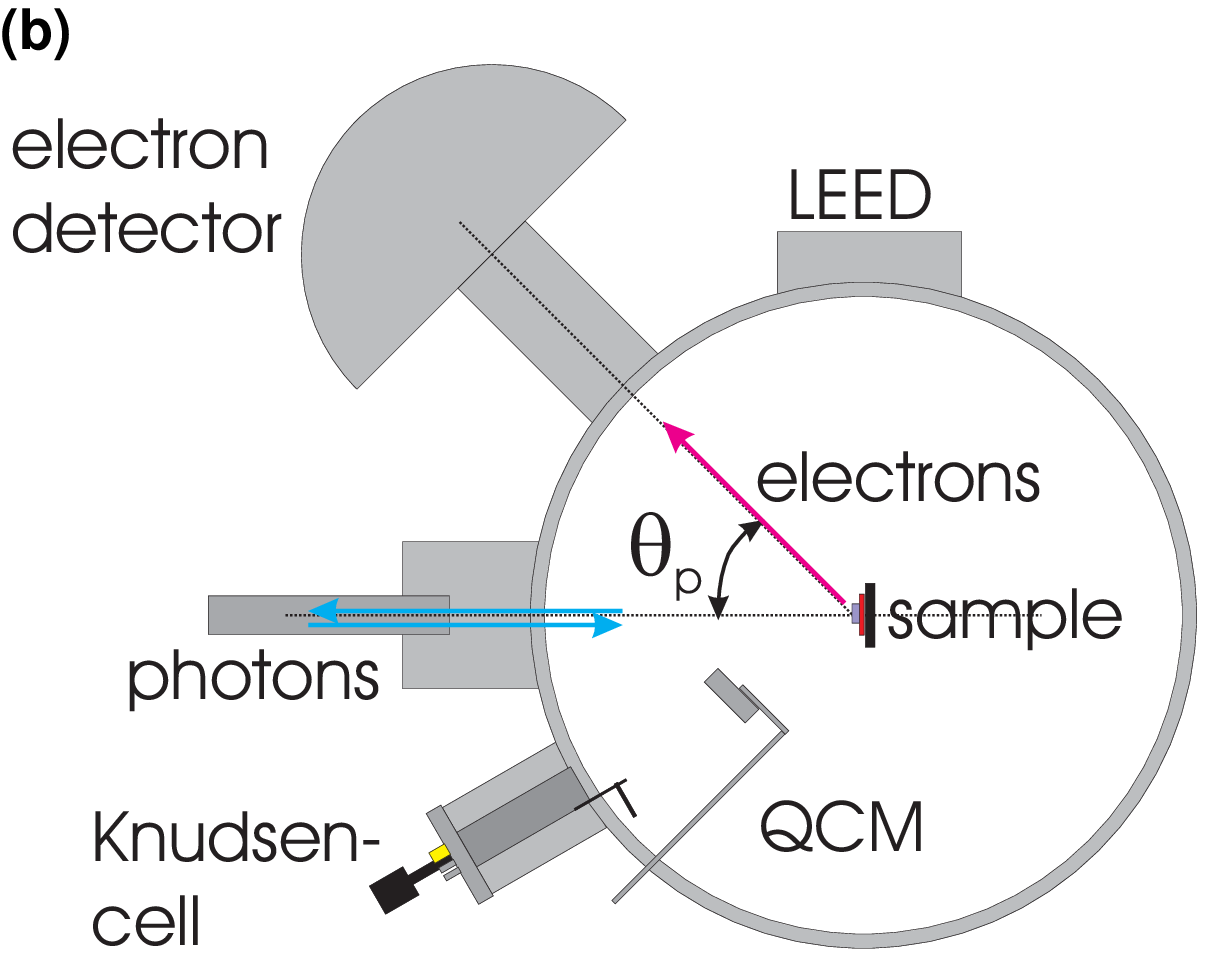} 
  \end{minipage}
  \caption{(Color online) (a) Molecular structure of PTCDA
    (3,4,9,10-perylene-tetracarboxylic-dianhydride) with chemically
    inequivalent carbon and oxygen atoms. (b) Experimental setup at the X-ray
    standing wave beamline ID32 (ESRF).}
  \label{fig:xswsetup}
\end{figure}

In this context the perylene derivative PTCDA
(3,4,9,10-perylene-tetracarboxylic-dianhydride, Fig.~\ref{fig:xswsetup}a) has
long been regarded as a model
system.\cite{jung_jmst93,gloeckler_ss98,fenter_prb97,krause_ass01,krause_epl04}
In particular, the adsorption of PTCDA on Ag(111) has been studied in detail
using different
techniques.\cite{krause_ass01,krause_epl04,tautz_prb02,kilian_ss04,zou_ss06}
After the average bonding distance of $\sim 2.85\,$\AA{} on silver had been
established by surface X-ray diffraction\cite{krause_jcp03}, X-ray standing
wave measurements\cite{hauschild_prl05} provided a refined result.  The
finding by Hauschild \textit{et al.}\cite{hauschild_prl05} that the adsorbed
molecule exhibits a significant and complex distortion has intensified the
interest of experimentalist and theoreticians alike. More recently it was
pointed out that the equilibrium distance of PTCDA on Ag(111) derived from
density functional calculations deviates notably, i.e.\ $0.55\,$\AA, from the
experimental bonding distance.\cite{rurali_prl05_comment} Regarding the
molecular distortion of PTCDA on Ag(111) theoretical results themselves are
still controversial.\cite{rurali_prl05_comment,hauschild_prl05_reply}

To broaden the experimental base and provide further much needed benchmarks
for calculations, we studied PTCDA monolayers on Cu(111) and Ag(111) using the
X-ray standing wave (XSW) technique. The model-free and precise bonding
distances, $d_H$, reported below show interesting patterns, most prominently a
non-trivial substrate dependence.

\section{Experimental Details}
\label{sec:experiment}

\subsection{Beamline setup and sample preparation}

X-ray standing wave experiments, which depend on the relatively weak
photoemission signals from organic adsorbates, require a brilliant and tunable
X-ray beam. Therefore, we performed our experiments at the undulator beamline
ID32 of the European Synchrotron Radiation Facility (ESRF).  Using the first
order back-reflections at $2.63\,$keV for Ag(111) and $2.98\,$keV for Cu(111)
we generated the X-ray standing wave field by Bragg reflection.  The
experimental end-station at ID32, an ultra-high vacuum chamber with a
hemispherical electron analyzer (energy resolution $\Delta E/E \sim 10^{-4}$),
was adapted for the preparation of organic thin films, see
Fig.~\ref{fig:xswsetup}b.

The single crystals, which were mounted on a variable-temperature
high-precision manipulator, were cleaned by repeated cycles of argon ion
bombardment. After annealing at $600\,$--$\,700\,$K we obtained suitable
surfaces as has been verified by X-ray photoelectron spectroscopy (XPS) and
low energy electron diffraction (LEED). We evaporated purified PTCDA at
typical rates of less than $1\,$ML/min with the substrate at $340\,$K,
monitoring the process with a quartz crystal microbalance close to the
substrate.  By heating the samples just below the desorption temperature of
the first layer we obtained well-ordered monolayers of PTCDA.

\subsection{Substrate characterization}

\begin{figure}[htbp]
  \centering 
  \includegraphics[width=8.6cm]{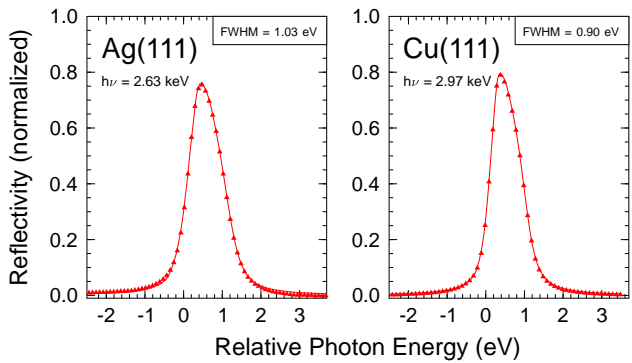}
  \caption{(Color online) Normal incidence reflectivity measurements around the
    first order Bragg reflection for Cu(111) and Ag(111). The solid line
    represents the reflectivity $R(E)$ calculated by dynamical diffraction
    theory with additional broadening due to the mosaicity of the sample and
    the finite monochromator resolution. The origin of the relative energy
    scale used throughout this article refers to the Bragg peak position as it
    would be observed without refraction inside the crystal.}
  \label{fig:darwin}
\end{figure}

The X-ray reflectivity around the substrate Bragg condition was measured at a
small angle relative to the incoming beam (Fig.~\ref{fig:darwin}).
Because noble metal crystals exhibit a certain mosaic spread that contributes
to the broadening of the Darwin curve, we checked the reflectivity signal to
identify suitable positions on the substrate.  Given the intrinsic width of
the Bragg reflections -- $0.96\,$eV for Ag(111) and $0.89\,$eV for Cu(111) --
the experimental reflection curves shown in Fig.~\ref{fig:darwin} illustrate
the crystal quality of the chosen surface positions.  A least-square fit to
the Bragg peaks using dynamical diffraction theory yields the effective
standing wave field, i.e.\ the reflectivity $R(E)$ and the phase $\nu(E)$
between the incoming and outgoing wave. Both quantities characterize the
substrate and enter directly into the XSW analysis.

\section{Results and Analysis}
\label{sec:results}

\subsection{Photoemission analysis}
\label{sec:results-a}

The core-level spectra provide essential information about the molecular and
electronic structure of the adsorbate system. Hence we briefly discuss the
relevant features in the monolayer and multilayer signals, focusing on the
aspects required for the interpretation of the XSW data.
\begin{figure}[htbp]
  \centering
  \includegraphics[width=8.6cm]{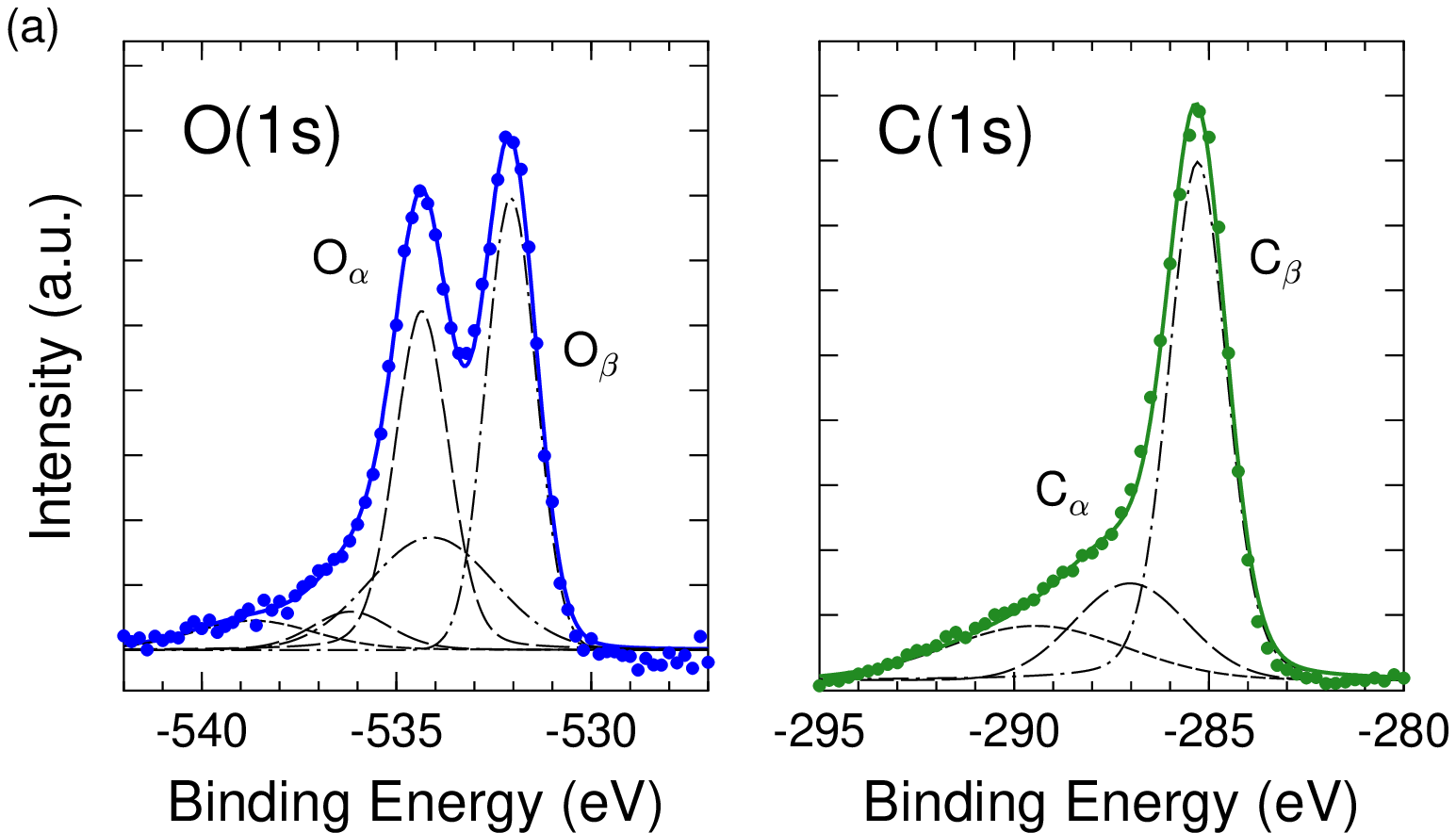}
  \includegraphics[width=8.6cm]{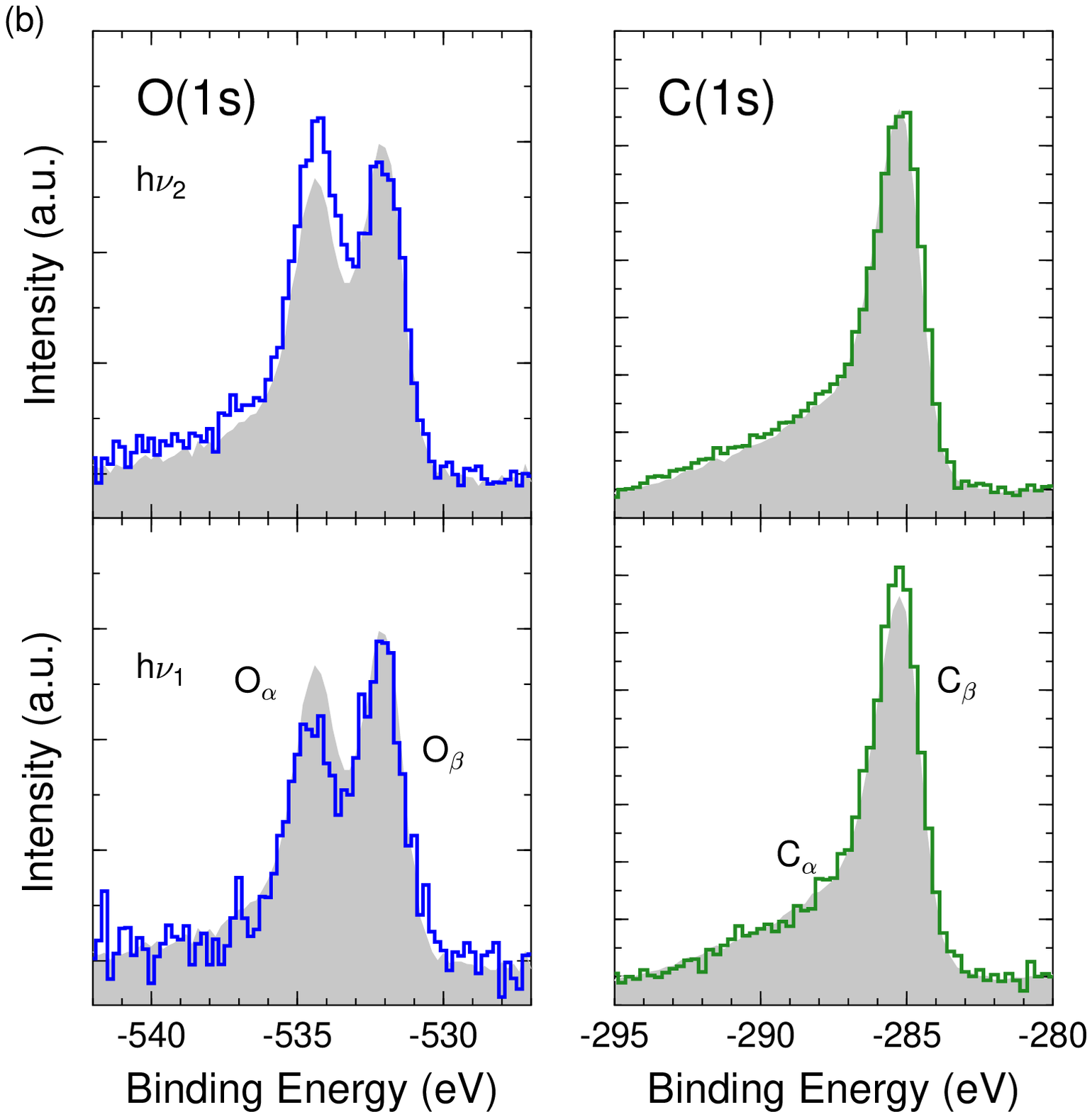}
  \caption{(Color online) Photoemission spectra from a monolayer of PTCDA on
    Cu(111). (a) The background corrected O(1s) and C(1s) sum spectra provide
    a reference signal, which can be described by chemically shifted main
    components $\alpha$ and $\beta$ plus shake-up states. (b) Comparison of
    spectra taken with different photon energies ($h\nu_1=2968.55$\,eV
    and $h\nu_2=2969.25$\,eV) with the shaded reference illustrate the
    different XSW characteristics in the O(1s) region.}
  \label{fig:xps_cu}
\end{figure}

\subsubsection{PTCDA monolayer}
\label{sec:results-a1}

As shown in previous experimental and theoretical studies both the carbon
C(1s) and oxygen O(1s) signal consist of several components, which can be
related to the molecular structure of PTCDA.\cite{schoell_jcp04} To obtain a
reference spectrum with a low noise level, we added up all background
corrected spectra in the XSW series (cf.\ Fig.~\ref{fig:xps_cu}a).  Because of
the limited energy resolution in the XSW setup a slightly simplified model
adequately explains our experimental line-shapes.

The principal component C$_\beta$ at a binding energy of $E_B=-285.1\,$eV in
the carbon core-level spectrum is related to excitations from the perylene
core of PTCDA.  The weaker signal C$_\alpha$ found at higher binding energies
originates from the carboxylic carbon atoms. Additional photoemission
intensity towards even higher energies can be related to shake-up processes
and inelastic background.  Similarly, the oxygen core-level spectrum allows to
distinguish the chemically inequivalent oxygen atoms in the molecule.  The
spectrum in Fig.~\ref{fig:xps_cu}a shows two main peaks at $E_B=-534.3\,$eV
and $E_B=-532.1\,$eV, which are associated with the anhydride (O$_\alpha$) and
carboxylic oxygen (O$_\beta$). To model the spectrum and preserve the
stoichiometric 2:1-ratio of both oxygen components, two corresponding shake-up
peaks were included in the analysis.

As illustrated in Fig.~\ref{fig:xps_cu}b the C(1s) signal shows a constant
shape throughout the XSW scan, whereas the relative intensity of the oxygen
components deviates significantly from the reference spectrum. This
observation allows two conclusions regarding the adsorption geometry on
Cu(111): First, the carbon core of PTCDA -- represented by the C$_\alpha$ and
C$_\beta$ components -- is planar within the experimental resolution.  Second,
the carboxylic and anhydride oxygen atoms are located at different bonding
distances $d_H$. The full XSW analysis verifying these statements follows
further below.

\subsubsection{PTCDA multilayer}
\label{sec:results-a2}

\begin{figure}[htbp]
  \centering 
  \includegraphics[width=8.6cm]{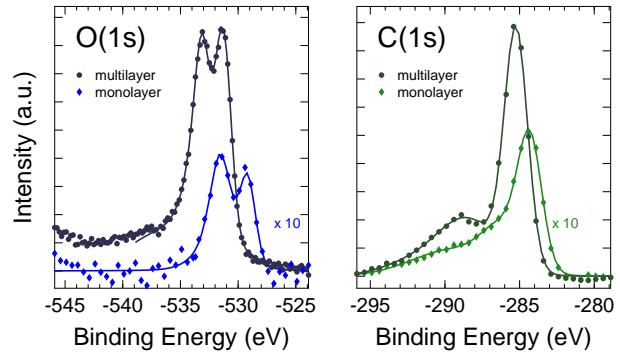}
  \caption{(Color online) Comparison of photoemission spectra
    on Ag(111) measured on a monolayer and a thick film of PTCDA. The
    significant core-level shifts are caused by the strong interaction of
    PTCDA with the substrate. For a more detailed discussion of the multilayer
    spectra see Ref.~\onlinecite{schoell_jcp04}.}
  \label{fig:xps_ag}
\end{figure}

The bonding distance of the first molecular layer is regarded as an important
factor influencing the growth of multilayer
films.\cite{krause_prb02,krause_ss04} Depending on the strength of the
adsorbate interaction thicker films can exhibit important changes in the
photoemission spectra.  Indeed, a comparison of the monolayer signal with
spectra taken on PTCDA multilayers shows significant core-level shifts both
for the C(1s) and O(1s) lines on Ag(111), see Fig.~\ref{fig:xps_ag}.

Obviously, the electronic structure within the first layer is strongly
affected by the bonding to the metal surface.\footnote{Further changes of the
  core-level signals observed in Fig.~\ref{fig:xps_ag}, e.g.\ a clearly
  resolved satellite peak in the C(1s) multilayer spectrum, are beyond the
  scope of this study. We refer to other more detailed
  studies\cite{schoell_jcp04} on the electronic structure of PTCDA
  multilayers.} The strong substrate-adsorbate interaction of PTCDA on
silver\cite{tautz_prb02,zou_ss06} -- presumably accompanied by a charge
transfer from the substrate -- should not least be evidenced by the bonding
distance.

\subsection{XSW analysis}
\label{sec:results-b}

\subsubsection{XSW fundamentals}
\label{sec:results-b1}

The normalized photoelectron yield $Y_p(\Omega)$ from the adsorbate atoms,
given by\cite{zegenhagen_ssr93,bedzyk_review,woodruff_rpp05}
\begin{equation}
  Y_p(\Omega ) = 1 + S_R R + 2 \sqrt{R} f_\mathit{eff}
   \cos (\nu - 2\pi P_\mathit{eff}),
  \label{eq:effec_XSWyield}
\end{equation}
depends sensitively on the (effective) coherent position $P_\mathit{eff}$ and
coherent fraction $f_\mathit{eff}$. These parameters contain all structural
information to be obtained from the coherently ordered monolayer. Following
the procedure described in Ref.~\onlinecite{gerlach_prb05} we model the
observed photoelectron yield with the previously determined reflectivity $R$
and phase $\nu$. A least-square fitting routine then finds the effective
parameters $P_\mathit{eff}$ and $f_\mathit{eff}$ associated with the
scattering atoms.

The first-order corrections to the dipole
approxi\-mation\cite{vartanyants_ssc00} included in
Eq.~(\ref{eq:effec_XSWyield}) are applied, in particular by measuring $S_R$ on
multilayers of PTCDA.\cite{schreiber_ssl01} Finally, we substitute the
effective parameters inserting\cite{gerlach_prb05}

\begin{equation}
  f_\mathit{eff} = |S_I| f_H \qquad \mathrm{and} \qquad P_\mathit{eff} =
  P_H - \psi / 2\pi
  \label{eq:non_dipole}  
\end{equation}
in Eq.~(\ref{eq:effec_XSWyield}). Using the well-established $|S_I|$- and
$\psi$-values in Tab.~\ref{tab:non_dipole} we deduce the coherent fraction,
$f_H$, and coherent position, $P_H$.  For molecules in a lying-down
configuration the phase $0\leq P_H \leq 1$ enters the ratio of the adsorbate
distance $d_H$ and the substrate lattice plane spacing $d_0$ according to
$d_H/d_0=1+P_H$.  From this equation we obtain model independent
results\cite{cheng_prl03} for the bonding distances $d_H$ of individual atomic
species.

\subsubsection{Bonding distances I}
\label{sec:results-b2}

\begin{figure}[htbp]
  \centering
  \begin{minipage}{\columnwidth} 
  \includegraphics[width=8.6cm]{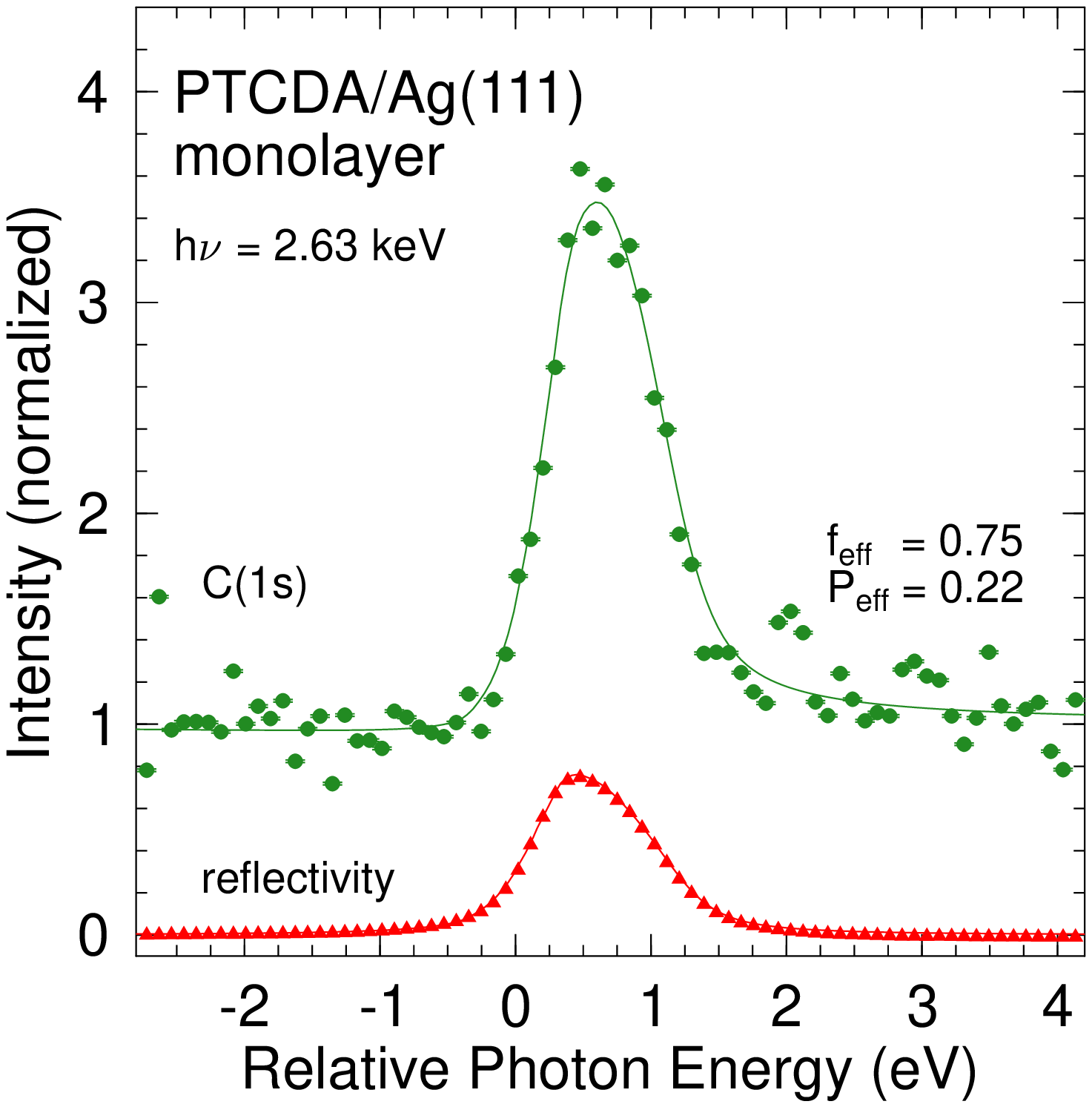}
  \hspace{2mm}
  \includegraphics[width=8.6cm]{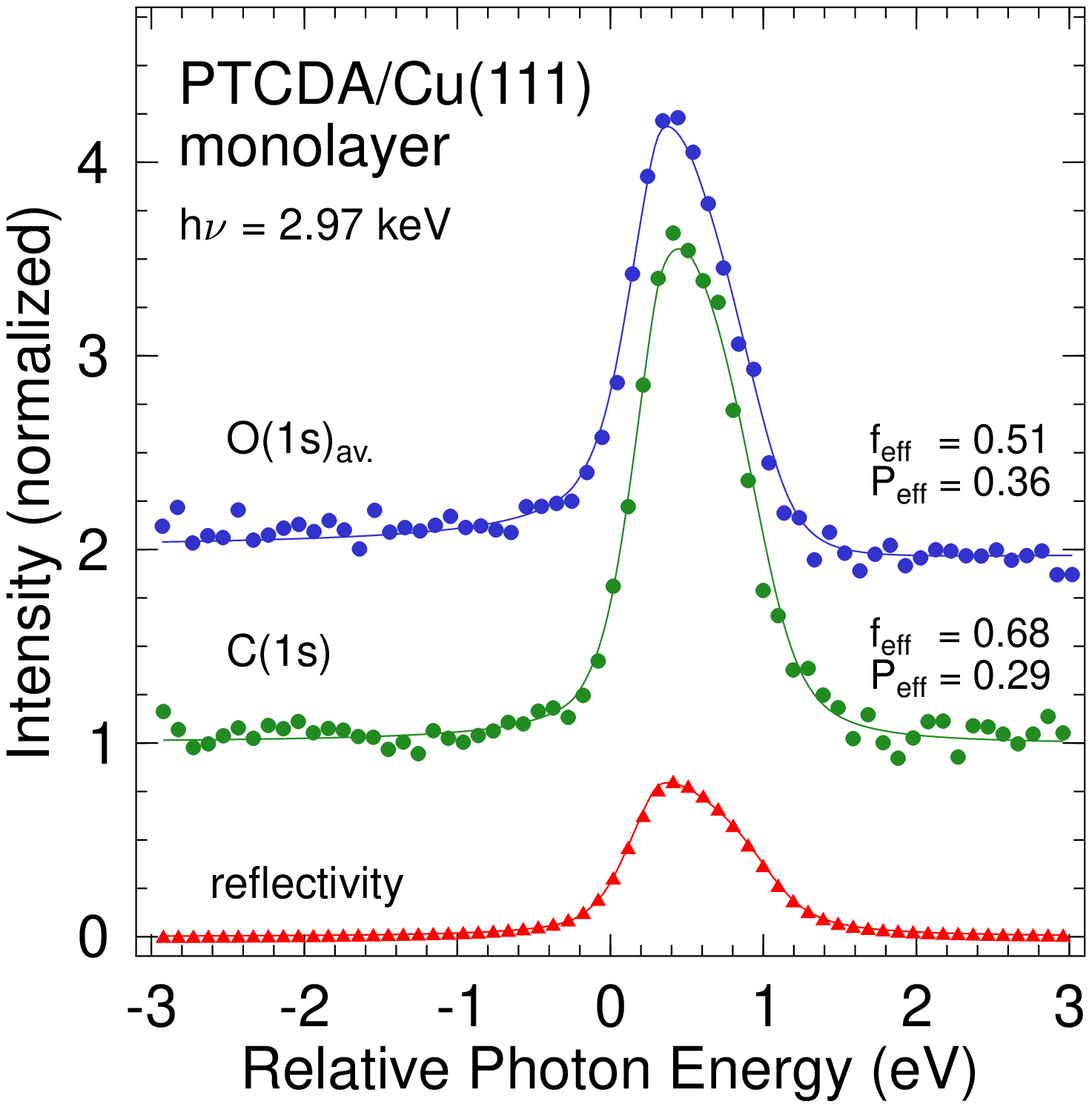}
  \end{minipage}
  \caption{(Color online) X-ray reflectivity and normalized
    photoemission yield data measured in back-reflection geometry on a
    monolayer of PTCDA on Ag(111) and Cu(111). The solid lines through the
    experimental XSW data $Y^\mathit{exp}$ show fit results $Y^\mathit{fit}$
    based on Eq.~(\ref{eq:effec_XSWyield}) with the corresponding coherent
    fraction $f_\mathit{eff}$ and coherent position $P_\mathit{eff}$. The
    oxygen dataset in the right panel is plotted with an offset of 1 for
    clarity.}
  \label{fig:cu111coh}
\end{figure}
The typical datasets presented in Fig.~\ref{fig:cu111coh} demonstrate that the
adsorption geometry of PTCDA yields standing wave characteristics with a
similar overall shape.  To determine the precise bonding distances and
possible distortions of the molecule therefore requires measurements with a
low noise level. Generally, we improve the statistics of our XSW data by
adding several scans, each one obtained by integration of the background
corrected and normalized photoelectron spectra. Importantly, the XSW yields
derived within this approach do not depend on particular assumptions about
line shapes and relative intensities of the components.  Because of the planar
carbon core this averaging procedure conserves the full information of the
C(1s) spectra and provides the XSW yield associated with the carbon atoms of
PTCDA.  Regarding the oxygen signal on the other hand we do not differentiate
between the carboxylic and anhydride oxygen and first derive an averaged
oxygen position.

The XSW characteristics of PTCDA on Ag(111) and Cu(111) shown in
Fig.~\ref{fig:cu111coh} were analyzed using Eq.~(\ref{eq:effec_XSWyield}). The
C(1s) data come with large coherent fractions indicating a fairly high order
in the monolayer: Least-square fits on silver (copper) yield
$f_\mathit{eff}=0.75$ ($f_\mathit{eff}=0.68$) and $P_\mathit{eff}=0.22$
($P_\mathit{eff}=0.29$) for the carbon core.  The corresponding oxygen result
on Cu(111), being $f_\mathit{eff}=0.51$ and $P_\mathit{eff}=0.36$, shows a
slightly reduced coherent fraction. For the oxygen result of PTCDA/Ag(111) we
refer to Ref.~\onlinecite{hauschild_prl05}.
\begin{table}[htbp]
  \centering
  \begin{ruledtabular}
    \begin{tabular}{l|D{.}{.}{1.6}|D{.}{.}{1.6}|D{.}{.}{1.6}|D{.}{.}{1.6}}
      & \multicolumn{2}{c|}{Cu(111)} & \multicolumn{2}{c}{Ag(111)} \\[1ex] 
      & \multicolumn{1}{c|}{C(1s)} & \multicolumn{1}{c|}{O(1s)} &
      \multicolumn{1}{c|}{C(1s)} &  \multicolumn{1}{c}{O(1s)} \\[1ex]
      \hline
      $S_R$    & 1.85(10) & 1.72(10)  & 1.89(5) & 1.89(5) \\
      $|S_I|$  & 1.43   & 1.36  & 1.45 & 1.45 \\
      $\psi$   & -0.055  & -0.075  & -0.066   & -0.093 \\
    \end{tabular}
  \end{ruledtabular}
  \caption{Non-dipolar parameters used in the XSW analysis: The $S_R$-values,
    which were measured on multilayer films, are in excellent agreement with
    previous results on PTCDA.\cite{schreiber_ssl01} The factor $|S_I|$ and
    the phase  $\psi$ were determined as described in
    Ref.~\onlinecite{gerlach_prb05}.}
  \label{tab:non_dipole}
\vspace{3ex}
%
%
  \begin{ruledtabular}
    \begin{tabular}{l|D{.}{.}{1.6}|D{.}{.}{1.6}|D{.}{.}{1.6}|D{.}{.}{1.6}|}
      & \multicolumn{2}{c|}{Cu(111)} & \multicolumn{2}{c}{Ag(111)} \\[1ex] 
      & \multicolumn{1}{c|}{C(1s)} &
      \multicolumn{1}{c|}{O(1s)$_\mathrm{av.}$} &
      \multicolumn{1}{c|}{C(1s)} &
      \multicolumn{1}{c}{O(1s)$_\mathrm{av.}$\footnote{Taken from Ref.~\onlinecite{hauschild_prl05}}} \\[1ex]
      \hline
      $f_\mathit{eff}$ & 0.68(6) & 0.51(6) & 0.75(12) &  \\
      $P_\mathit{eff}$ & 0.29(1) & 0.36(2) & 0.22(2)  &  \\
      $f_H$            & 0.48(4) & 0.37(4) & 0.52(8) & 0.57   \\
      $P_H$            & 0.28(1) & 0.35(2) & 0.21(2)  & 0.18    \\[1ex] 
      $d_H$            & 2.66(2)\,\textnormal{\AA} & 2.81(3)\,\textnormal{\AA} &
      2.86(5)\,\textnormal{\AA} & 2.78\,\textnormal{\AA}
    \end{tabular}
  \end{ruledtabular}
  \caption{XSW results for a monolayer of PTCDA on Cu(111) and Ag(111): The
      effective parameters are obtained from the data in
      Fig.~\ref{fig:cu111coh}.  From the  coherent position $P_H$ we derive
      the atomic position $d_H$ relative to the Bragg planes of the
      substrate. The statistical uncertainties noted in parentheses follow
      from the confidence analysis in Fig.~\ref{fig:conf_level}.}
  \label{tab:xsw_results}
\end{table}

Applying the non-dipolar corrections of Eq.~(\ref{eq:non_dipole}) we calculate
the corresponding atomic positions $d_H$ as described above -- neglecting a
possible small relaxation of the outer substrate layers. For the carbon core
we find $d_H = 2.86 \pm 0.05\,$\AA{} on Ag(111) and $d_H = 2.66 \pm
0.02\,$\AA{} on Cu(111).  Remarkably, the oxygen atoms of PTCDA on copper
reside at an averaged position of $d_H = 2.81\pm 0.03\,$\AA, i.e.\ 
$0.15\,$\AA{} \emph{above\/} the central perylene core of the molecule.  All
structural parameters for PTCDA on Cu(111) and Ag(111) obtained so far are
collected in Tab.~\ref{tab:xsw_results}.

\begin{figure}[tbp]
  \centering 
  \includegraphics[width=8.6cm]{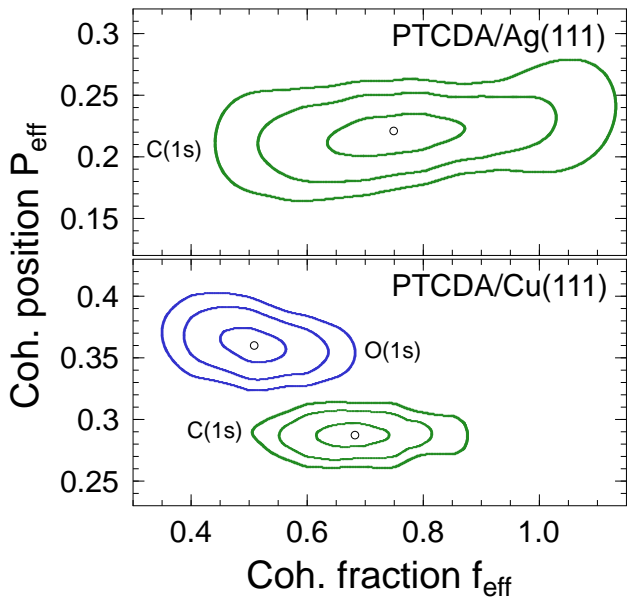}
  \caption{(Color online) $\chi^2$-contour map showing the statistical
    uncertainties in $f_\mathit{eff}$ and $P_\mathit{eff}$ for the
    least-square fits presented in Fig.~\ref{fig:cu111coh}. The different
    contour lines give the $68$\%-, $95$\%-, and $99.7$\%-confidence
    levels around the global $\chi^2$-minimum.}
  \label{fig:conf_level}
\end{figure}

To derive meaningful error bars for $d_H$ and see whether the molecular
distortion is statistically significant, we performed a detailed confidence
analysis. For the XSW datasets $Y_i^\mathit{exp}$ shown in
Fig.~\ref{fig:cu111coh} we sampled the parameter space and calculated the
$\chi^2$-values in the vicinity of the obtained minima.
Fig.~\ref{fig:conf_level} shows the corresponding confidence levels in the
$(f_\mathit{eff}, P_\mathit{eff})$-plane with constant
\begin{equation}
  \label{eq:chisquare}
  \chi^2_{f_\mathit{eff},P_\mathit{eff}} = \sum_i 
  \left\{(Y_i^\mathit{exp} -  Y_i^\mathit{fit} )/\sigma_i \right\}^2.
\end{equation}

The errors $\sigma_i$ of the XSW yield, that are entering as weighting factors
in the $\chi^2$-calculation, were derived from the counting statistics of the
photoelectron spectra.\footnote{Since the signal-to-background ratio of the
  photoemission spectra can be rather low, the energy dependence of $\sigma_i$
  roughly follows the substrate standing wave characteristic.  Hence the
  statistical weight of the data points in the XSW scans varies strongly.} We
observe well-defined minima for all datasets with an uncertainty depending on
the noise in the XSW scans, see Fig.~\ref{fig:conf_level}.  Generally, the
errors of the coherent fraction are larger than the corresponding errors of
the coherent position, i.e.\ $\Delta f_\mathit{eff}/f_\mathit{eff} > \Delta
P_\mathit{eff}/P_\mathit{eff}$.  Hence the atomic positions can be determined
quite precisely and for our data we infer $\Delta d_H = 0.02\,$\AA{}
($0.03\,$\AA) for carbon (oxygen) on copper and $\Delta d_H = 0.05\,$\AA{} for
carbon on silver. Moreover, the results for PTCDA/Cu(111) in
Fig.~\ref{fig:conf_level} demonstrate that our finding with the average oxygen
position being above the carbon is valid beyond a $99.7$\%-confidence level.

\subsubsection{Bonding distances II}

\begin{figure}[htbp]
  \centering
  \includegraphics[height=8.6cm,angle=270]{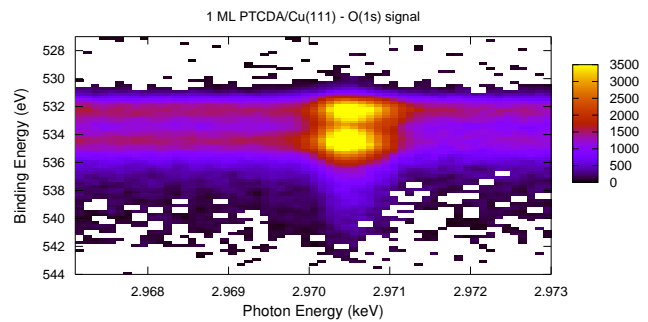}
  \caption{(Color online) Intensity map of the O(1s) core-level spectra
    measured on PTCDA/Cu(111).  The anhydride and carboxylic oxygen atoms --
    with the O$_\alpha$-component at $E_B=-534.3\,$eV) and O$_\beta$ at
    $E_B=-532.1\,$eV -- exhibit different XSW characteristics.}
  \label{fig:intensity_map_cu111}
\end{figure}

The different distances of both oxygen species in PTCDA/Cu(111) are
illustrated by the intensity map in Fig.~\ref{fig:intensity_map_cu111}, where
relative XPS intensities can be compared. While the carboxylic oxygen
O$_\beta$ shows a fairly symmetric XSW characteristic, the anhydride oxygen
O$_\alpha$ exhibits a rapidly decreasing photoemission yield on the high
energy tail of the Bragg peak. This difference reveals that the O$_\alpha$
component has a higher coherent position and is further away from the
substrate than the O$_\beta$ atoms.

\begin{figure}[htbp]
  \centering 
  \includegraphics[width=8.6cm]{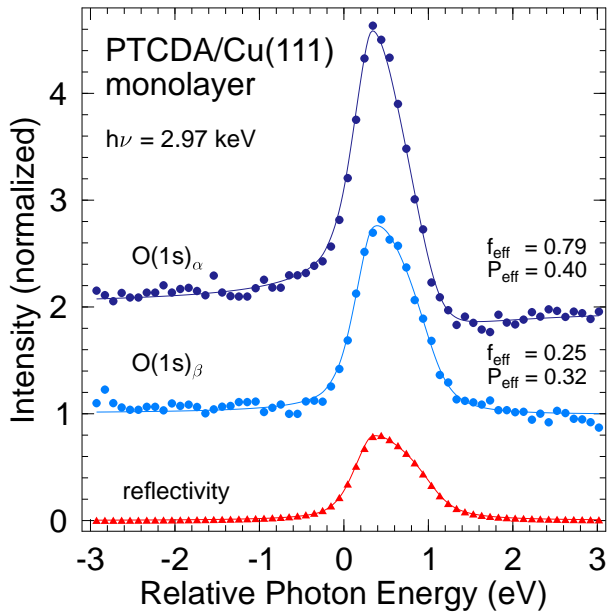}
  \caption{(Color online) X-ray standing wave scans on a monolayer of
    PTCDA on Cu(111). Comparison of XSW signals from different oxygen species
    O$_\alpha$ and O$_\beta$. An effective coherent fraction of 0.25 for
    O$_\beta$ in comparison to 0.79 for O$_\alpha$ and 0.68 for C might
    indicate the presence of more than one O$_\beta$ site with respect to the
    substrate crystal lattice. The separation of the overlapping photoemission
    intensities, however, also introduces a sizeable uncertainty of
    $f_\mathit{eff}$ that should be taken into consideration.}
  \label{fig:cu111coh_ox}
\end{figure}

Beyond this qualitative consideration a refined analysis of the photoemission
spectra can provide the exact positions of both oxygen species.  Yet,
separating the O$_\alpha$ and O$_\beta$ component is not trivial as strong
shake-up states in the core-level spectra complicate the
procedure.\cite{hauschild_prl05} Accordingly, we started with the O(1s) sum
spectrum to build a model which reproduces the data (cf.\ 
Fig.~\ref{fig:xps_cu}a).  Good agreement with the experiment was obtained
using two chemically shifted main lines -- each one with a corresponding
shake-up peak $1.95\,$eV below -- and an additional shake-off peak towards
higher binding energies. The individual spectra in the XSW series were
analyzed using this model, with the peak amplitudes being the only free
parameters. Keeping the relative intensities of the shake-up peak and the main
line constant we obtain the XSW yield of the oxygen components in
PTCDA/Cu(111). From these separate datasets we derive the coherent positions
of $P_\mathit{eff}=0.32$ for the carboxylic and $P_\mathit{eff}=0.40$ for the
anhydride oxygen, see Fig.~\ref{fig:cu111coh_ox} for details.
The corresponding bonding distances confirm the distortion of PTCDA on Cu(111)
with $d_H = 2.89\,$\AA{} for anhydride oxygen and $d_H = 2.73\,$\AA{} for
carboxylic oxygen.  This splitting of $0.16\,$\AA{} around the averaged oxygen
distance $d_H = 2.81\,$\AA{} depends marginally on the fitting model with an
estimated error of $0.06\,$\AA.

\section{Discussion}
\label{sec:discussion}

The bonding distances found on Ag(111) and Cu(111) are much smaller than the
molecular stacking distance $d_{(102)}=3.21\,$\AA{} measured in PTCDA single
crystals.  This observation conveys the relatively strong bonding of the
molecule to these substrates. A more detailed inspection, however, reveals
some remarkable differences between the adsorption of PTCDA on copper and
silver as well as between the adsorption of PTCDA and other $\pi$-conjugated
molecules like F$_{16}$CuPc.\cite{gerlach_prb05}

\begin{figure}[htbp]
  \centering
    \includegraphics[width=8.6cm]{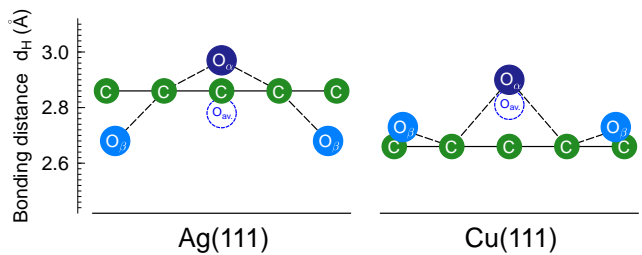}
  \caption{(Color online) Different adsorption geometries of PTCDA on Ag(111)
    and Cu(111) measured by XSW. On copper both oxygen species O$_\alpha$ and
    O$_\beta$ -- distinguishable due to a chemical shift in the photoemission
    spectra -- are located above the carbon plane. On silver, however, the
    carboxylic oxygen atoms O$_\beta$ are bent towards the
    surface.\cite{hauschild_prl05} }
  \label{fig:adsorp_geom_ptcda}
\end{figure}

\paragraph{PTCDA/Ag(111) --}
In agreement with previous XSW results\cite{hauschild_prl05} we measured a
carbon distance of $d_H = 2.86\,$\AA{} on Ag(111).  Interestingly, this value
coincides with the bonding distance derived from the electron density profile
of PTCDA multilayer films on Ag(111)\cite{krause_jcp03}, indicating that the
distance of the first layer is not markedly affected by the presence (growth)
of further layers.  As reported in Ref.~\onlinecite{hauschild_prl05} the
average oxygen position is \textit{below\/} the carbon core at $d_H
=2.78\,$\AA.  A significant splitting of $0.29\,$\AA{} --~with the anhydride
oxygen O$_\alpha$ above ($d_H =2.97\,$\AA) and the carboxylic oxygen O$_\beta$
below ($d_H =2.68\,$\AA) the carbon plane~-- is
observed\cite{hauschild_prl05}, see Fig.~\ref{fig:adsorp_geom_ptcda}.

\paragraph{PTCDA/Cu(111) --}
We found a carbon distance of $d_H = 2.66\,$\AA{} on Cu(111), i.e.\ a value
that is smaller than on silver, but very similar to the result of F$_{16}$CuPc
on copper. In contrast to Ag(111) the average oxygen position on copper is
\textit{above\/} the carbon core at $d_H = 2.81\,$\AA. While the oxygen
splitting is qualitatively similar --~with the O$_\alpha$ further from
substrate than the O$_\beta$ component~-- here both species are $0.07\,$\AA{}
and $0.23\,$\AA, respectively, above the carbon plane. Thus, we find a
splitting of $0.16\,$\AA{} on copper which is only half the value observed on
Ag(111).

\paragraph{PTCDA/Au(111) --}
X-ray standing wave studies of PTCDA on Au(111) reported
recently\cite{sokolowski_dpg06} gave a carbon distance of $d_H = 3.34\,$\AA.
Again, this bonding distance agrees with X-ray reflectivity data taken on
multilayers of PTCDA/Au(111)\cite{fenter_prb97} which implied positions around
$3.35\,$\AA. Thus PTCDA molecules adsorb at comparatively large distances on
gold, suggesting a weaker interaction with the substrate. Similar findings
made with high-resolution electron energy-loss spectroscopy
(HREELS)\cite{eremtchenko_njp04} and scanning tunneling microscopy
(STM)\cite{nicoara_oe06} measurements on Au(111) support this conclusion.

\paragraph{PTCDA vs.\ F$_{16}$CuPc --}  
Most prominently, we find that the substrate dependence of $d_H$ is much
smaller for PTCDA than for F$_{16}$CuPc. While for PTCDA the carbon distances
on Ag(111) and Cu(111) differ by only $0.20\,$\AA, the corresponding
difference for F$_{16}$CuPc is as large as $0.64\,$\AA. Moreover, in
F$_{16}$CuPc monolayers the fluorine atoms experience an upward bending both
on Cu(111) and Ag(111), whereas PTCDA molecules exhibit a molecular distortion
that depends on the substrate.

The different molecular distortions of PTCDA on Ag(111) and Cu(111) shown in
Fig.~\ref{fig:adsorp_geom_ptcda} cannot easily be explained without extensive
theoretical work. The observed differences and similarities on these
substrates, however, indicate that the bonding of the molecule to the metal
occurs mainly through its carbon core. While the oxygen atoms reside at
similar distances relative to the copper and silver substrate, the carbon core
is noticeably closer to the Cu(111) surface -- as might be expected for the
bonding to the smaller copper atoms.  Obviously, the different distortion of
the molecule should affect the charge distribution within the adsorbate.
Indeed, photoemission spectra measured on the valence bands of PTCDA/Ag(111)
and PTCDA/Cu(111)\cite{duhm_unpublished} point towards remarkable differences
in the electronic structure.

\section{Summary and Conclusions}
\label{sec:conclusion}

By XSW measurements of PTCDA on Cu(111) and Ag(111) we show that the bonding
distance and the adsorption geometry, i.e.\ the bending of the oxygen atoms,
depends in a non-trivial way on the substrate.

We hope that our results will stimulate further theoretical work in this area.
Calculations on the adsorbate structure of large molecules would greatly
promote our understanding of these systems and could also provide new insight
in the electronic properties of the organic-inorganic interface.

\section*{Acknowledgments}

The authors gratefully acknowledge the ESRF for providing excellent
facilities, and thank J.\ Pflaum for purifying the PTCDA material.  This work
was financially supported by the DFG.


\end{document}